\documentclass[sn-vancouver]{sn-jnl}

\theoremstyle{thmstyleone}%
\theoremstyle{thmstyletwo}%

\theoremstyle{thmstylethree}%

\usepackage{multirow}
\usepackage[strings]{underscore}
\usepackage{amsmath}
\usepackage{booktabs}

\begin{document}

\title[Large-Kernel Attention]{Large-Kernel Attention for 3D Medical Image Segmentation}


\author[1,2]{\fnm{Hao} \sur{Li}}\email{hao.li19@imperial.ac.uk}

\author[1]{\fnm{Yang} \sur{Nan}}\email{y.nan20@imperial.ac.uk}

\author[3,4]{\fnm{Javier} \sur{Del Ser}}\email{javier.delser@tecnalia.com}


\author*[1,5]{\fnm{Guang} \sur{Yang}}\email{g.yang@imperial.ac.uk}


\affil[1]{\orgdiv{National Heart and Lung Institute, Faculty of Medicine}, \orgname{Imperial College London}, \city{London}, \country{United Kingdom}}

\affil[2]{\orgdiv{Department of Bioengineering, Faculty of Engineering}, \orgname{Imperial College London}, \city{London}, \country{United Kingdom}}

\affil[3]{\orgname{TECNALIA, Basque Research \& Technology Alliance (BRTA)}, \city{Derio}, \country{Spain}}

\affil[4]{\orgname{University of the Basque Country (UPV/EHU)}, \city{Bilbao}, \country{Spain}}

\affil[5]{\orgname{Royal Brompton Hospital}, \city{London}, \country{United Kingdom}}


\abstract{Automatic segmentation of multiple organs and tumors from 3D medical images such as magnetic resonance imaging (MRI) and computed tomography (CT) scans using deep learning methods can aid in diagnosing and treating cancer. However, organs often overlap and are complexly connected, characterized by extensive anatomical variation and low contrast. In addition, the diversity of tumor shape, location, and appearance, coupled with the dominance of background voxels, makes accurate 3D medical image segmentation difficult. In this paper, a novel large-kernel (LK) attention module is proposed to address these problems to achieve accurate multi-organ segmentation and tumor segmentation. The advantages of convolution and self-attention are combined in the proposed LK attention module, including local contextual information, long-range dependence, and channel adaptation. The module also decomposes the LK convolution to optimize the computational cost and can be easily incorporated into FCNs such as U-Net. Comprehensive ablation experiments demonstrated the feasibility of convolutional decomposition and explored the most efficient and effective network design. Among them, the best Mid-type LK attention-based U-Net network was evaluated on CT-ORG and BraTS 2020 datasets, achieving state-of-the-art segmentation performance. The performance improvement due to the proposed LK attention module was also statistically validated.}

\maketitle

\section{Introduction}\label{sec1}

Malignant tumors and other organ illnesses have long been a problem for humans, seriously endangering their lives and general well-being. Worldwide, millions of people die from cancer each year, making it the leading cause of mortality \cite{sungGlobalCancerStatistics2021}. Nevertheless, early identification and therapy are still the most effective means of enhancing cancer survival. Identifying the location of organs and lesions is a crucial step in the diagnostic process and plays a vital role in treating diseases. In general, locating organs and lesions from medical images such as Magnetic Resonance Imaging (MRI) and Computed Tomography (CT) is a segmentation task. Clinicians can determine the location, size, and subtype of a tumor through the precise segmentation of tumors. This benefits not only in the diagnostic process but also in planning radiation therapy or surgery. On the other hand, accurate organ segmentation can help clinicians select personalized treatment strategies for various patients, enabling the practice of precision medicine and individualized care, which can lessen the patient's financial and psychological burdens. Additionally, the segmentation of longitudinal MRI images can be utilized to track tumor development or shrinkage as well as the response and recovery of diseased organs to therapeutic interventions. Therefore, the research and implementation of medical image segmentation are of major significance.

Segmentation of organs and lesions is typically performed manually by experienced radiologists in current clinical practice. Observing medical images to differentiate human organs, tissues, and lesions is a challenging and time-consuming endeavor. Additionally, because manual labeling results rely heavily on the radiologist's expertise and subjective judgment, they are rarely reproducible and might even involve human bias. Consequently, these problems contribute to the low practicability of manual segmentation. Automated or computer-aided segmentation approaches can solve these issues by requiring less labor and producing objective, reproducible results for later disease diagnosis and management. As a result, automated medical picture segmentation has been thoroughly researched and has emerged as the benchmark.

With the increase in GPU computing power and the quick advancement of deep learning technology in recent years, the field of image semantic segmentation has also grown rapidly. Natural image segmentation approaches based on fully convolutional neural networks (FCN) \cite{longFullyConvolutionalNetworks2015} have been more developed over time. In the meantime, medical image segmentation remains a formidable challenge, as medical images are characterized by uneven grayscale, significant contrast variation, and substantial noise. Since U-Net \cite{ronnebergerUNetConvolutionalNetworks2015} was published, medical image semantic segmentation has also undergone tremendous development. Numerous convolutional neural network variations, such as FCN and U-Net, are rapidly being applied.

However, the existing technology for the automatic segmentation of medical images lacks sufficient intelligence and precision. Organs are challenging to differentiate because of their overlap and intricate connections. Additionally, the wide variations in anatomical structure and the low contrast make segmentation tasks ambiguous. Moreover, for lesion segmentation, tumors can arise in any organ location and exhibit a wide range of size, shape, and appearance \cite{soltaninejadAutomatedBrainTumour2017}. Second, in many cases, the tumor volume is rather small relative to the entire scan, resulting in the dominance of the background noise \cite{dsouzaMRITumorSegmentation2018}. All of these issues lower the segmentation accuracy. In clinical practice, even minute inaccuracies in medical picture segmentation might result in misdiagnosis. Therefore, segmentation models based on deep learning have significant space for development in this discipline.

Long-range self-attention can be used to enable the network to learn only the truly crucial information \cite{jiaH2NFNetBrainTumor2021}, such as the organ boundaries or tumor-related features. It is a mechanism for adaptive selection based on the inputs' features. Different self-attention techniques have been used in medical image segmentation \cite{chenTransUNetTransformersMake2021, sinhaMultiScaleSelfGuidedAttention2021,valanarasuMedicalTransformerGated2021}. They have obtained superior performance compared to traditional FCNs because of their efficiency in capturing long-range dependencies. Despite recent attempts \cite{chenTransUNetTransformersMake2021, sinhaMultiScaleSelfGuidedAttention2021,valanarasuMedicalTransformerGated2021}, self-attention has several shortcomings when it comes to medical image segmentation since it was designed for Natural Language Processing (NLP). First, it analyzes images as one-dimensional sequences, ignoring the structural details required for obtaining morphological features in medical images. Second, since 3D scans like MRI or CT are too computationally expensive with quadratic complexity, most self-attention research is 2D-based. Third, it disregards the necessity of channel adaptation for attention processes. For image semantic segmentation tasks, different channels usually represent features of different objects. Thus, adaptation in channel maps is important for attentions to build dependencies within channels \cite{sinhaMultiScaleSelfGuidedAttention2021,guoAttentionMechanismsComputer2022,huGatherExciteExploitingFeature2018,parkBAMBottleneckAttention2018,wangResidualAttentionNetwork2017,wooCBAMConvolutionalBlock2018}.

In order to address these issues, this paper introduces a novel large-kernel (LK) attention module for enhancing medical image segmentation. The LK attention module combines self-attention and convolution's advantages, such as long-range dependence, spatial adaptation, and local contextual information, and avoiding their disadvantages, such as disregarding channel adaptation and computational complexity. This paper is based on our previous work on MRI brain tumor segmentation at the Medical Image Understanding and Analysis Conference (MIUA) \cite{LKAU-Net}. On this basis, we optimized the LK attention model, conducted comprehensive ablation experiments to demonstrate its feasibility and explore more efficient design and deployment strategies. We also further investigated whether LK attention could improve the performance of CT multi-organ segmentation to expand the application scope and adaptability of LK attention in medical imaging and segmentation tasks. The following highlights the key contributions of this paper:

\begin{itemize}
    \item A novel LK attention module utilizing decomposed LK convolutions was proposed, which combines the advantages of convolution and self-attention while avoiding their disadvantages.
    \item A U-Net architecture that efficiently incorporated LK attention was proposed for the segmentation of 3D medical images. By adaptively amplifying the weights of key features while reducing the weights of noisy voxels, the LK attention-based U-Net can accurately identify the locations of various organs and tumor subregions.
    \item In publicly available datasets for evaluating multi-organ and tumor segmentation, LK attention-based U-Net outperformed state-of-the-art methods in delineating all targets.
    \item Extensive ablation experiments were performed, and the findings validated the effectiveness of the decomposition of the LK convolution and investigated the optimal deployment and design strategies for the LK attention module.
    \item The proposed LK attention module is easy to integrate into any other neural network. Quantitative studies demonstrated that it could effectively improve the accuracy of medical image segmentation and provide local explanations.
\end{itemize}

The rest of the article is structured as follows: Section 2 will briefly review related work. Section 3 will detail our segmentation method, including the LK attention module and network architecture. Section 4 will illustrate the experimental setup, and results and discussion will be presented in Section 5. The conclusion will be given in the final Section 6.

\section{Related Work}

In this section, we will briefly review the recent work related to multi-organ segmentation and tumor segmentation, including some applications of self-attention.

\subsection{Multi-organ Segmentation}

Multi-organ segmentation, which comprehensively classifies voxels into multiple organ classes rather than just organs or other tissues, gives a broader viewpoint on the task of organ segmentation. This involves identifying which organ type a particular voxel belongs to, in addition to determining if it belongs to an organ \cite{wangAutomaticMultiorganSegmentation2014}. Due to the increased data volume and image complexity, the automatic segmentation of multiple organs in 3D medical images is a tedious challenge. 

A method for segmenting 3D CT images using the majority voting was proposed in \cite{belagiannisDeepLearningData2016} based on the FCN. In \cite{dou3DDeeplySupervised2017}, a neural network dubbed 3D DSN avoids unnecessary computation and overfitting via volume-to-volume learning, making it suited for applying to cardiac and hepatic anatomy. H.R. Roth et al.\cite{rothApplicationCascaded3D2018} presented a coarse-to-fine method for multi-organ segmentation that included two stages. The 3D FCN in the first stage extracts candidate regions coarsely, whereas the second 3D FCN focuses on potential organ region boundaries in a cascaded way, hence minimizing the number of voxels to be processed. Similar research was conducted by \cite{chenAutomaticAbdominalMultiOrgan2017} employing cascaded 3D FCNs for dual-energy CT. \cite{kakeya3DUJAPANetMixture2018} presented a 3D-U-JAPA-net based on transfer learning, whereas \cite{zhouSemiSupervisedMultiOrganSegmentation2018} created a semi-supervised network to fully exploit the unlabeled data. To save GPU memory, \cite{tangSpatialContextAwareSelfAttention2021} suggested combining 2D and 3D models, performing segmentation using 2D convolutions, and extracting spatial information from 3D models. 

Although FCNs have been shown to be very successful, learning long-range spatial relationships is challenging due to the localization of convolutional layers. The UNETR architecture was proposed by \cite{hatamizadehUNETRTransformers3D2022}, who was inspired by transformers used in NLP. The transformer acting as an encoder enables U-Net to gather global information and long-range spatial relationships, leading to superior segmentation results.

\subsection{Tumor Segmentation}

Identification of tumors can be aided by image analysis across various imaging modalities. The Brain Tumor Segmentation Challenge (BraTS) compiles a well-known public multi-modal MRI dataset. The BraTS challenge compares cutting-edge brain tumor segmentation methods annually \cite{bakasAdvancingCancerGenome2017,bakasIdentifyingBestMachine2019,menzeMultimodalBrainTumor2015}. T1-weighted (T1), post-contrast T1-weighted (T1ce), T2-weighted (T2), and T2 fluid attenuated inversion recovery (FLAIR) 3D MRI modalities are available for each patient case.

Since 2014, deep learning algorithms have been extensively researched for tumor segmentation in the BraTS challenge \cite{guan3DAGSEVNetAutomatic2022,huangDeepMultiTaskLearning2021,isenseeNnUNetBrainTumor2021,jiaH2NFNetBrainTumor2021,jiangTwoStageCascadedUNet2020,myronenko3DMRIBrain2019,soltaninejadAutomatedBrainTumour2017,soltaninejadSupervisedLearningBased2018,wangModalityPairingLearningBrain2021,zhangMENetMulti2021}. Myronenko \cite{myronenko3DMRIBrain2019} won the BraTS 2018 competition by training an asymmetrical U-Net with a broader encoder and an additional variational decoder branch that provided further regularization. A two-stage cascaded asymmetrical U-Net comparable to Myronenko \cite{myronenko3DMRIBrain2019} was proposed by Jiang et al. \cite{jiangTwoStageCascadedUNet2020}. The first step generated a coarse prediction, whereas the second stage utilized a larger network to refine the outcome. In order to automatically adapt the traditional U-Net to a particular dataset with just minor alterations, Isensee et al. \cite{isenseeNnUNetBrainTumor2021} adopted a self-configuring framework called nnU-Net. Wang et al. \cite{wangModalityPairingLearningBrain2021} suggested a modality-pairing learning method that uses the layer connection on parallel branches to extract the complicated interactions and rich information between various MRI modalities. Jia et al. \cite{jiaH2NFNetBrainTumor2021} created the Hybrid High-resolution and Non-local Feature Network (H2NF-Net), which used parallel multi-scale convolutional blocks to utilize multi-scale features and preserve high-resolution features representation simultaneously. The self-attention mechanism implemented in this study permits the aggregation of local information across spatial locations and the acquisition of long-range dependence.

\section{Method}

Our method is detailed in this section, including the new LK attention module and the modified U-Net based on the LK attention module for 3D medical image segmentation. 

\subsection{LK Attention}

\begin{figure}[h]
\includegraphics[width=\textwidth]{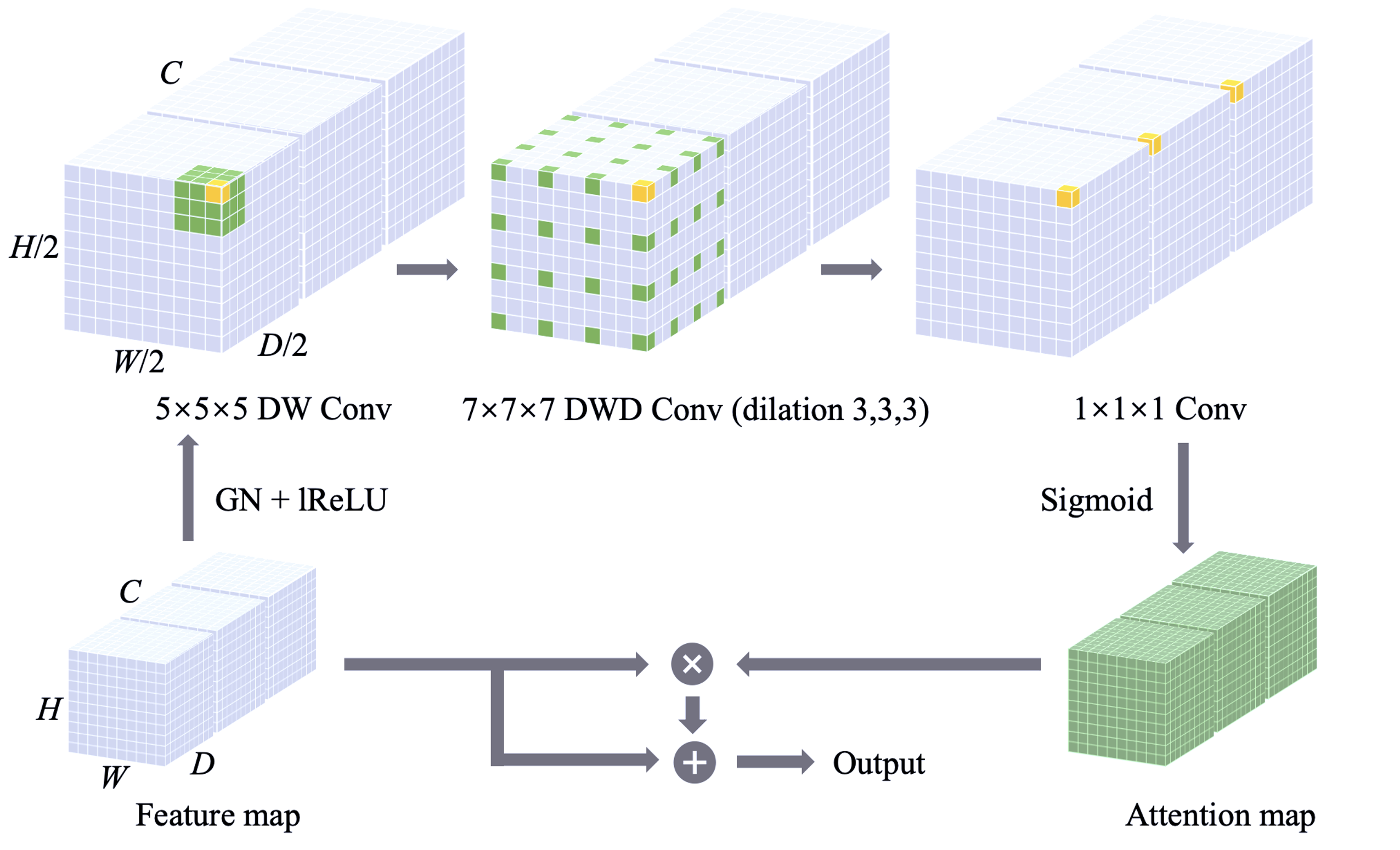}
\caption{LK attention module. The decomposed LK convolution is applied on the feature map after group normalization (GN) and leaky ReLU (lReLU). The attention map is obtained by sigmoid activation, which is then multiplied and summed elementwise with the original feature map to generate the module output. The figure shows a representative decomposition of a $21\times21\times21$ convolution into a $5\times5\times5$ depth-wise (DW) convolution, a $7\times7\times7$ depth-wise dilated (DWD) convolution with dilation of 3, and a $1\times1\times1$ convolution. The position of the kernel is indicated by colored voxels, and the yellow voxels show the kernel's centers. (The figure only illustrates a corner of the feature space of the decomposed LK convolution and disregards the zero-padding.)} 
\label{fig1}
\end{figure}

Numerous studies have demonstrated that the integration of diverse attention mechanisms has the potential to enhance segmentation performance. The attention map reflects the relative significance across the feature space, which necessarily involves the capture of correlations between various locations. The self-attention can be used to discover long-range dependence, but it has several disadvantages, as stated in the previous section. Applying large-kernel convolution to establish long-distance relationships and generate the attention map is an alternative method \cite{guoAttentionMechanismsComputer2022,guoVisualAttentionNetwork2022,huGatherExciteExploitingFeature2018,parkBAMBottleneckAttention2018,wangResidualAttentionNetwork2017,wooCBAMConvolutionalBlock2018}. Nevertheless, this strategy substantially increases the computational cost.

To address these limitations and maximize the benefits of self-attention and large-kernel (LK) convolution, we developed an LK attention module (shown in Figure~\ref{fig1}). Assuming $K$ is the number of channels, a $K\times{K} \times{K}$ LK convolution was decomposed into a $(2d-1)\times(2d-1)\times(2d-1)$ depth-wise (DW) convolution, a $\frac{K}{d}\times\frac{K}{d}\times\frac{K}{d}$ depth-wise dilated (DWD Conv) convolution with dilation of d and a $1\times1\times1$ convolution. For an input with dimensions of $H\times W\times D\times C$, the number of parameters (${\text{N}}_{\text{PRM}}$) and the number of floating-point operations (FLOPs) for the original LK convolution and its decomposition can be calculated as follows:
\begin{align}
{\text{N}}_{\text{PRM,O}}&=C\times(C\times(K\times K\times K)+1), \\
{\text{FLOPs}}_\text{O}&=C\times(C\times(K\times K\times K)+1)\times H\times W\times D,
\end{align}
\begin{align}
\label{eqn3}
\begin{split}
    {\text{N}}_{\text{PRM,D}}{}&=C\times((2d-1)\times(2d-1)\times(2d-1)\\
    &~~~+\frac{K}{d}\times\frac{K}{d}\times\frac{K}{d}+C+3),
\end{split}\\
\begin{split}
    {\text{FLOPs}}_\text{D}{}&=C\times((2d-1)\times(2d-1)\times(2d-1)\\
    &~~~+\frac{K}{d}\times\frac{K}{d}\times\frac{K}{d}+C+3)\times H\times W\times D,
\end{split}
\end{align}
where O and D represent the original LK convolution and decomposed LK convolution, respectively. To determine the optimal d such that $N_{PRM}$ is minimal for a particular kernel size K, we set the first derivative of Equation~\ref{eqn3} to 0 and then solved as follows:
\begin{equation}\label{eq5}
    \frac{d}{dd^\ast}\left(C\left(\left(2d^\ast-1\right)^3+\left(\frac{K}{d^\ast}\right)^3+C+3\right)\right)=0,
\end{equation}\label{eq6}
\begin{equation}
    24d^2-24d-\frac{3K^3}{d^4}+6=0.
\end{equation}

In Equation~\ref{eq5}, the superscript $\ast$ distinguishes dilation $d$ from derivation $d$. For $K=21$, solving Equation~\ref{eq6} numerically yielded an optimal approximation of $d$ of approximately 3.4159. As shown in Table~\ref{tab1}, the number of parameters can be significantly lowered with a dilation rate of 3. We can also observe that as the number of channels increases, the decomposition becomes more efficient.

\begin{table}[h]
\begin{center}
\begin{minipage}{\textwidth}
\caption{Complexity analysis: comparison of the number of parameters $N_{PRM}$ for a $21\times21\times21$ convolution.}
\label{tab1}
\begin{center}
\begin{tabular}{rrrr}
\toprule
\multicolumn{1}{l}{$C$} & \multicolumn{1}{l}{${\text{N}}_{\text{PRM,O}}$} & \multicolumn{1}{l}{${\text{N}}_{\text{PRM,D}}$} & \multicolumn{1}{l}{${\text{N}}_{\text{PRM,D}}/{\text{N}}_{\text{PRM,O}}$}\\
\midrule
32	& 9.48 M	    & 16.10 k	& 0.17\%\\
64	& 37.94 M	& 34.24 k	& 0.09\%\\
128	& 151.75 M	& 76.67 k	& 0.05\%\\
256	& 606.99 M	& 186.11 k	& 0.03\%\\
512	& 2427.98 M	& 503.30 k	& 0.02\%\\
\bottomrule
\end{tabular}
\end{center}
\footnotetext[]{The subscripts $O$ and $D$ denote the original convolution and the proposed decomposed convolution, respectively. $C$: number of channels.}
\end{minipage}
\end{center}
\end{table}

The entire LK attention module is formulated as follows:
\begin{equation}
    A=\sigma_{\text{sigmoid}}\left({\text{Conv}}_{1\times1\times1}\left({\text{Conv}}_{\text{DW}}\left({\text{Conv}}_{\text{DWD}}\left(\sigma_{\text{lReLU}}\left(\text{GN}\left(Input\right)\right)\right)\right)\right)\right)\\,
\end{equation}
\begin{equation}
    Output=A\otimes\left(\sigma_{\text{lReLU}}\left(\text{GN}\left(Input\right)\right)\right)+\sigma_{\text{lReLU}}\left(\text{GN}\left(Input\right)\right),
\end{equation}
where $A$ denotes the attention map, and GN is the group normalization. $\sigma_{\text{lReLU}}$ and $\sigma_{\text{sigmoid}}$ denote to leaky ReLU activation function and sigmoid activation function, respectively. The LK Attention module's output is formed by multiplying and summing the input feature map and the attention map element by element. Using the LK attention module, we can extract long-range correlations within a feature space and generate the attention map with minimal computing complexity and parameters.

\subsection{LK Attention-based U-Net}

\begin{figure}[h]
\includegraphics[width=\textwidth]{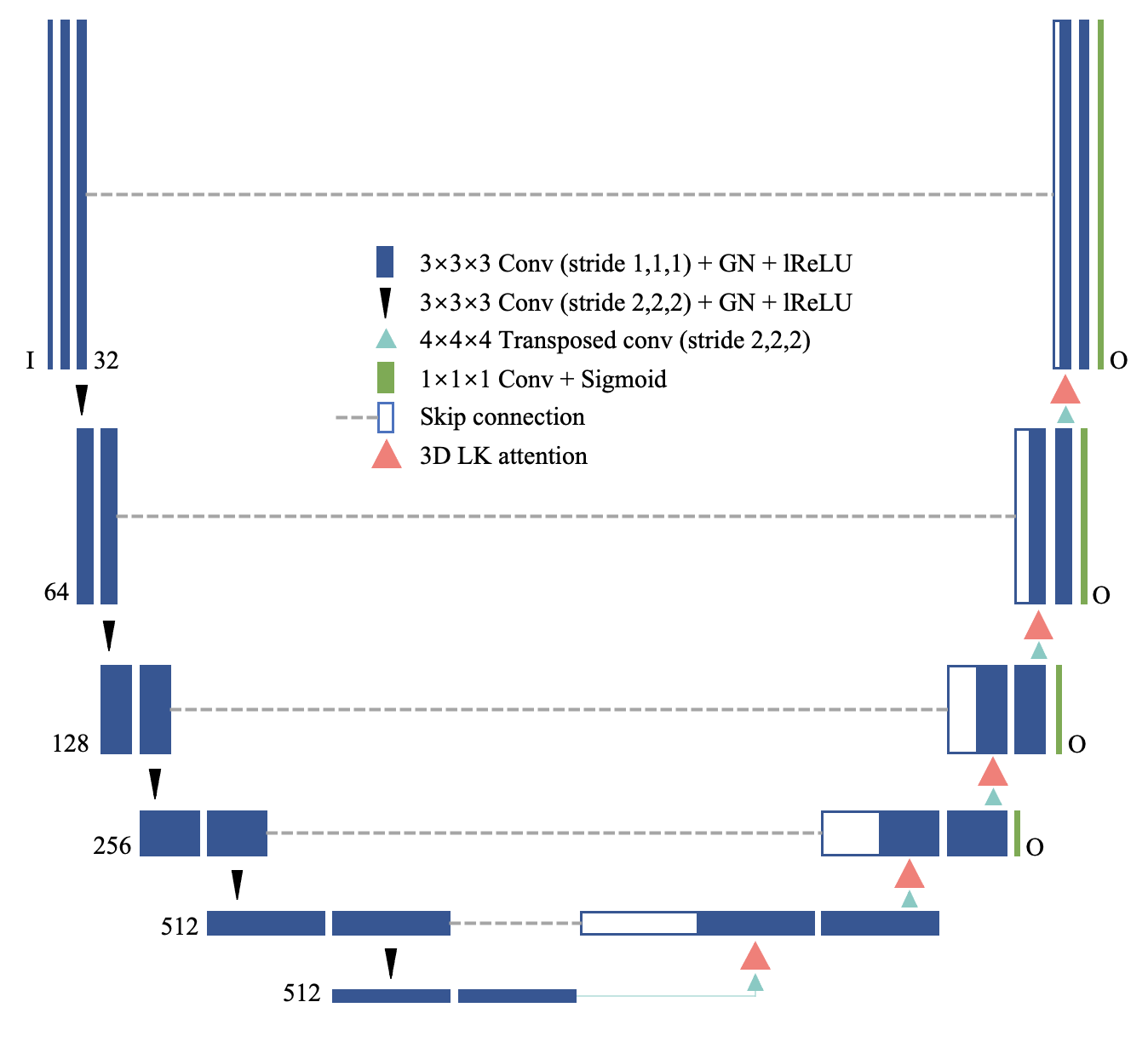}
\caption{The network architecture of our proposed LK attention-based U-Net.} 
\label{fig2}
\end{figure}

The U-Net \cite{ronnebergerUNetConvolutionalNetworks2015} has served as a basis for numerous studies on medical image processing. Its capacity to capture fine object features utilizing skip-architecture is particularly advantageous for precise segmentation. As shown in Figure~\ref{fig2}, the 3D LK attention-based U-Net architecture is based on the U-Net and comprises an encoding path of feature extraction and a decoding path of inference with the skip connection. 

\subsubsection{Encoder}

The encoder is composed of convolution blocks of six scales. Each block contains two convolution layers with a $3\times3\times3$ kernel, GN, and lReLU (with a slope of 0.01). The input data of $I$ channels is convoluted by 32 kernels to generate initial 32 feature maps, and the channel number $I$ corresponds to the number of imaging modalities. Between the two scales, a stride-2 $3\times3\times3$ convolution is used to downsample the feature map by 2 and increase the number of channels to a maximum of 512. The deepest feature map is $1/32$ of the original size.

\subsubsection{LK Attention-Based Decoder} 

The architecture of the decoder is identical to that of the encoder, using $4\times4\times4$ transposed convolution for upsampling. The LK attention module can be applied to each upsampled feature map to form a fully applied (Full) network as in our previous paper. The details of the LK attention module for the Full network are shown in Table~\ref{tab2}. At the last layer, a $1\times1\times1$ convolution is applied to compress the channel number O according to the number of segmentation classes, followed by the softmax/sigmoid to generate probability maps for different organs or tumor regions. Additional softmax/sigmoid outputs were added to all scales except the two lowest levels for deep supervision and boost gradient propagation. 

\begin{table}[h]
\footnotesize
\centering
\caption{Details of LK attention modules in the Full LK attention-based U-Net}
\label{tab2}
\begin{tabular}{lllllll}
\toprule
\multirow{2}{*}{\textbf{Scale}}	& \multicolumn{2}{l}{\textbf{DW Conv}} & \multicolumn{3}{l}{\textbf{DWD Conv}} & \textbf{Equal LK Conv}\\
& Kernel & Padding & Kernel & Dilation &Padding & Kernel\\
\midrule
10×12×8	&(3, 3, 3)	&(1, 1, 1)	&(3, 3, 3)	&(2, 2, 2)	&(2, 2, 2)	&(6, 6, 6)\\
20×24×16	&(3, 3, 3)	&(1, 1, 1)	&(3, 3, 3)	&(2, 2, 2)	&(2, 2, 2)	&(6, 6, 6)\\
40×48×32	&(3, 3, 3)	&(1, 1, 1)	&(5, 5, 5)	&(2, 2, 2)	&(4, 4, 4)	&(10, 10, 10)\\
80×96×64	&(5, 5, 5)	&(2, 2, 2)	&(5, 5, 5)	&(3, 3, 3)	&(6, 6, 6)	&(15, 15, 15)\\
160×192×128	&(5, 5, 5)	&(2, 2, 2)	&(7, 7, 7)	&(3, 3, 3)	&(9, 9, 9)	&(21, 21, 21)\\
\bottomrule
\end{tabular}
\end{table}

\section{Experiment}

The LK attention is evaluated on standard benchmarks: CT-ORG \cite{risterCTORGNewDataset2020} for multi-organ segmentation and BraTS 2020 for tumor segmentation. We first conducted extensive ablation experiments to evaluate the proposed module's effectiveness thoroughly.

\subsection{Data Acquisition}
The CT-ORG \cite{risterCTORGNewDataset2020} dataset consists of 140 CT images of six organ classes, including liver, lungs, bladder, kidneys, bones, and brain. Of the total 140 image volumes, 131 were dedicated CTs, and 9 were CT components collected during PET-CT examinations. Each image was acquired from a different patient. Most images displayed benign or malignant liver lesions; some showed metastasis from breast, colon, bone, and lung cancers. The images were collected from a variety of sources, including low-dose, high-dose, contrast, and non-contrast CT, with dedicated CTs ranging from 0.56 mm to 1 mm in axial resolution. With the aid of ITK-SNAP and morphological segmentation, manual segmentation was conducted on the test dataset (21 cases). Some images were received from the Liver Tumor Segmentation Challenge (LiTS) \cite{bilicLiverTumorSegmentation2019}.

The BraTS 2020 dataset was collected using various clinical protocols and scanners from different institutions. The ground truth (GT) labels are annotated by one to four raters and approved by specialists, which include the GD-enhancing tumor (ET), peritumoral edema (ED), and necrotic and non-enhancing tumor core (NCR + NET). The segmentation results are evaluated on three subregions of the tumor: the GD-enhancing tumor (ET), the tumor core (TC = ET + NCR + NET), and the whole tumor (WT = ET + NCR + NET + ED). The image modalities T1, T1ce, T2, and T2-FLAIR are co-registered to the same template with an image size of $240\times240\times155$. Afterward, they are interpolated to the same resolution ($1~\text{mm}^3$) and skull-stripped. Annotations are only available for the training set (369 cases). The evaluation of the independent validation set (125 cases) should be conducted on the official online platform (CBICA’s IPP\footnote{CBICA's Image Processing Portal (https://ipp.cbica.upenn.edu)}). Details of the two datasets are summarized in Table~\ref{tab3}.

\begin{table}[h]
\footnotesize
\centering
\caption{Details of datasets.}
\label{tab3}
\begin{tabular}{llllll}
\toprule
\textbf{Dataset} & \textbf{Modality}  & \textbf{Labels} & \textbf{Classes} & \textbf{Training set} & \textbf{Test set} \\
\midrule
CT-ORG           & CT                 & Organs          & 6                & 119                   & 21                \\
BraTS 2020       & MRI (4 modalities) & Brain tumors    & 3                & 369                   & 125              \\
\bottomrule
\end{tabular}
\end{table}

\subsection{Pre-processing and Data Augmentation}

For the CT-ORG dataset, our network takes an image volume of $128\times128\times256$ as input. To reduce GPU memory usage, all image volumes were resampled to $3~\text{mm}^3$. Resampling uses Gaussian smoothing to avoid aliasing artifacts, followed by resolution interpolation. All image volumes for the BraTS 2020 dataset are cropped to $160\times192\times128$ to reduce computational waste on background voxels. All input volumes are then pre-processed by intensity normalization.

Various data augmentation techniques have been applied to artificially increase dataset size and minimize the risk of overfitting. All augmentations are applied on-the-fly throughout the training to expand the training dataset indefinitely. Furthermore, to increase the variability of the generated data, all augmentations are applied randomly based on preset probabilities, and most parameters are also drawn randomly (see Table~\ref{tab4} for details).

\begin{table}[h]
\centering
\caption{Details of data augmentation strategies.}
\label{tab4}
\begin{tabular}{lll}
\toprule
\textbf{Methods}   & \textbf{Probability} & \textbf{Range}\\
\midrule
Brightness         & 30\%                 & $U(0.7, 1.3)$            \\
Contrast           & 15\%                 & $U(0.6, 1.4)$             \\
Gaussian Noise     & 15\%                 & variance $\sigma\sim U (0, 1)$          \\
Gaussian Blur      & 20\%                 & kernal $\sigma\sim U (0.5, 1.5)$         \\
Gamma Augmentation & 15\%                 & $\gamma\sim U(0.7, 1.5)$   
\\
Scaling            & 30\%                 & $U(0.65, 1.6)$            \\
Rotation           & 30\%                 & $U(-30, 30)$              \\
Elastic Transform  & 30\%                 & $\alpha\sim U(5, 10), \sigma=3\alpha$ \\
Flipping           & 50\%                 & along all axes\\
\bottomrule
\end{tabular}
\end{table}

\subsection{Training and Optimization}

The LK attention-based U-Net is trained separately on CT-ORG and BraTS 2020 training datasets. For the CT-ORG training set (119 cases), the network parameters are optimized for weighted soft Dice loss. The weight for each segmentation class is one minus the ratio of foreground voxels to background voxels. For the BraTS 2020 training set (369 cases), binary cross-entropy (BCE) and soft Dice losses are utilized. 

The adaptive moment estimator (Adam) optimizer was applied to optimize the parameters of the network. Each training process had 200 epochs with a batch size of 1 and an initial learning rate of 0.0003. All experiments were implemented with Pytorch 1.10 on an NVIDIA GeForce RTX 3090 GPU of 24GB VRAM.

\subsection{Evaluation Metrics}
The segmentation results were evaluated using the Dice score and 95 percent Hausdorff distance (HD95), which are defined as:
\begin{align}
    \text{Dice} &= \frac{2\lvert \mathcal{X} \cap \mathcal{Y} \rvert}{\lvert \mathcal{X} \rvert+\lvert \mathcal{Y}\rvert},\\
    \text{HD95} &= P_{95}\left(\max\left(\max_{x\in \mathcal{X} }{\min_{y\in \mathcal{Y}}{\lvert y-x\rvert}},\max_{y\in \mathcal{Y}}{\min_{x\in \mathcal{X} }{\lvert x-y\rvert}}\right)\right),
\end{align}
where $\mathcal{X}$ and $\mathcal{Y}$ are sets of GT and prediction, and P represents the percentile. HD95 indicates the 95th percentile of maximum distances between two boundaries, whereas the Dice score measures spatial overlap between the segmentation result and the GT annotation. The final performance of LK attention-based U-Net wasis evaluated using independent test sets from CT-ORG (21 cases) and BraTS 2020 (125 cases), respectively. The brain class was excluded from evaluation because it was present in only 8 of 119 training images, so the model had difficulty learning to differentiate it.

\section{Results and Discussion}

This section will first experimentally demonstrate the effectiveness of our LK attention module design, and then quantitatively analyze the segmentation results. The limitations of the proposed method will be also discussed in the last subsection.

\subsection{Qualitative Analysis of Ablation Experiments}

For the ablation study, the CT-ORG test dataset was used for evaluation, and the network without any attention module was adopted as the base model. We first verify the effectiveness of LK convolutional decomposition and then look for efficient ways to compute the attention map through different model variants.

We conducted ablation experiments by adding different single attention modules to the base network. By comparing the attention module using the original LK convolution with the attention module using the decomposed LK convolution, the decomposition of the LK convolution was proven to be effective and efficient. The comparative results in Table~\ref{tab5} show that the segmentation results of the two attention modules were very close at both the deepest and shallowest levels. For the bottleneck LK attention module, the segmentation of the decomposed LK convolution performed slightly worse than the original (average 0.09 difference in Dice score). Moreover, the segmentation performance of the decomposed LK convolution at the highest level was even better. The changes in Dice score were verified by paired t-tests in the test set, giving p-values of 0.094 and 0.122, respectively. In addition, we can also see that the decomposition of LK convolution significantly reduced the number of added parameters to about 0.5\% and 0.2\% of the original, respectively.

\begin{table}[h]
\setlength{\tabcolsep}{5pt}
\footnotesize
\centering
\caption{Quantitative results to compare the decomposed (D) 3D LK convolution with the original (O) 3D LK convolution.}
\label{tab5}
\begin{tabular}{llrllllll}
\toprule
\multirow{2}{*}{\textbf{Scale}} & \multirow{2}{*}{\textbf{LK Conv}} & \multirow{2}{*}{\textbf{$\textbf{N}_{\textbf{PRM}}$ (k)}} & \textbf{Dice$\uparrow$} &  &  &  &  &  \\
            &            &           & liver & bladder & lungs & kidneys & bone  & mean  \\ 
            \midrule
None (Base) & N/A        & 101017.22 & 95.81 & 86.81   & 94.23 & 92.11   & 88.20 & 91.43 \\
10×12×8     & O   & +56623.62 & 95.12 & 85.78   & 94.25 & 90.39   & 88.05 & 90.72 \\
10×12×8     & D & +291.33   & 95.00 & 85.75   & 93.97 & 90.38   & 88.04 & 90.63 \\
160×192×128 & O   & +9483.30  & 95.48 & 86.19   & 95.88 & 90.41   & 88.03 & 91.20 \\
160×192×128 & D & +16.10    & 95.55 & 86.23   & 95.77 & 90.59   & 88.09 & 91.25 \\
\bottomrule
\end{tabular}
\end{table}

The LK attention module can be applied to each upsampled feature map. However, the additional computational cost of a fully applied (Full) network is high, and the efficiency of its design deserves to be analyzed. Therefore, we explored many variants of attention modules with different sizes and positions, as shown in Table~\ref{tab6}. Applying decomposed LK attention modules with different kernel sizes at the same location ($160\times192\times128$) indicated that larger kernel coverage leads to better segmentation performance. Kernel coverage refers to the ratio of the kernel size to the feature space size. This is reasonable because convolutions with larger kernels capture correlations across longer distances more effectively. While decomposed LK convolutions with the same kernel size (6, 6, 6) at different locations show that the LK attention module worked best in the middle of the decoder. We can see that when the fixed kernel size LK attention module was applied to larger scales, its segmentation performance initially increased but started to decrease slightly due to the significant reduction of kernel coverage at high levels. Therefore, to balance the effects of kernel size and position, we applied the largest LK attention module in the middle, which achieved the highest Dice score. Therefore, the network structure utilizing LK attention in the middle of the decoder (Mid) is the most effective and efficient, with the number of added parameters being nearly one-sixth of the Full network.

\begin{table} [!h]
\setlength{\tabcolsep}{2pt}
\footnotesize
\centering
\caption{Quantitative results to compare 3D LK attention modules of different convolutional kernel sizes at different locations in the network.}
\label{tab6}
\begin{tabular}{llrrcccccc}
\toprule
\multirow{2}{*}{\textbf{Scale}} &
  \multirow{2}{*}{\textbf{\begin{tabular}[c]{@{}l@{}}Equal LK\\ Conv\end{tabular}}} &
  \multirow{2}{*}{\textbf{\begin{tabular}[c]{@{}l@{}}Kernel\\ Coverage\end{tabular}}} &
  \multirow{2}{*}{\textbf{NPRM (k)}} &
  \textbf{Dice$\uparrow$} &
   & & & & \\ & & & &
   \multicolumn{1}{l}{liver} & \multicolumn{1}{l}{bladder} & \multicolumn{1}{l}{lungs} & \multicolumn{1}{l}{kidneys} & \multicolumn{1}{l}{bone} & \multicolumn{1}{l}{mean}  \\
\midrule
None (Base) & N/A          & N/A     & 101017.22 & 95.81 & 86.81   & 94.23 & 92.11   & 88.20 & 91.43 \\
10×12×8     & (6, 6, 6)    & 22.50\% & +291.33   & 95.00 & 85.75   & 93.97 & 90.38   & 88.04 & 90.63 \\
20×24×16    & (6, 6, 6)    & 2.81\%  & +80.13    & 94.95 & 86.22   & 94.18 & 90.66   & 88.18 & 90.84 \\
40×48×32    & (6, 6, 6)    & 0.35\%  & +23.68    & 95.66 & 86.25   & 94.75 & 91.70   & 88.26 & 91.32 \\
80×96×64    & (6, 6, 6)    & 0.04\%  & +7.74     & 95.59 & 85.88   & 95.18 & 90.74   & 87.69 & 91.02 \\
160×192×128 & (6, 6, 6)    & 0.01\%  & +2.85     & 95.44 & 85.06   & 94.84 & 90.55   & 87.18 & 90.61 \\
160×192×128 & (10, 10, 10) & 0.03\%  & +5.98     & 95.58 & 85.32   & 95.14 & 90.68   & 87.40 & 90.83 \\
160×192×128 & (15, 15, 15) & 0.09\%  & +9.12     & 95.92 & 85.22   & 95.68 & 90.90   & 87.91 & 91.13 \\
160×192×128 & (21, 21, 21) & 0.24\%  & +16.10    & 95.55 & 86.23   & 95.77 & 90.59   & 88.09 & 91.25 \\
40×48×32    & (21, 21, 21) & 15.07\% & +76.67    & 96.12 & 86.48   & 97.40 & 92.26   & 88.51 & 92.15 \\
All (Full)  & see Table~\ref{tab2}  & N/A     & +444.06   & 96.12 & 86.63   & 95.56 & 91.70   & 88.45 & 91.69 \\
\bottomrule
\end{tabular}
\end{table} 

\subsection{Quantitative Analysis of Segmentation}

The evaluation of the segmentation performance of the proposed methods was conducted and compared with state-of-the-art methods, including CBAM \cite{wooCBAMConvolutionalBlock2018} using an independent CT-ORG test set (21 cases) and BraTS 2020 validation set (125 cases), which are shown in Table~\ref{tab7} and~\ref{tab8}. 

\begin{table} [!h]
\setlength{\tabcolsep}{1pt}
\footnotesize
\centering
\caption{Quantitative results of proposed methods compared to state-of-the-art methods for CT-ORG. (Bold numbers are the best results)}
\label{tab7}
\begin{tabular}{lcccccclcccccc}
\toprule
\multirow{2}{*}{\textbf{Method}} & \multicolumn{6}{l}{\textbf{Dice$\uparrow$}} && \multicolumn{6}{l}{\textbf{HD95$\downarrow$}} \\
 & \multicolumn{1}{l}{liver} & \multicolumn{1}{l}{bladder} & \multicolumn{1}{l}{lungs} & \multicolumn{1}{l}{kidneys} & \multicolumn{1}{l}{bone} & \multicolumn{1}{l}{mean} && \multicolumn{1}{l}{liver} & \multicolumn{1}{l}{bladder} & \multicolumn{1}{l}{lungs} & \multicolumn{1}{l}{kidneys} & \multicolumn{1}{l}{bone} & \multicolumn{1}{l}{mean} \\
 \midrule
U-Net \cite{ronnebergerUNetConvolutionalNetworks2015} & 94.83 & 76.79          & 93.85 & 89.35   & 85.43 & 88.05 && 2.71  & 4.64    & 14.10 & 4.87    & 6.27 & 6.52 \\
nnU-Net \cite{isenseeNnUNetBrainTumor2021}   & 95.48 & 86.00          & 95.21 & 91.74   & 87.84 & 91.25 && 1.81  & 3.02    & 9.67  & 3.11    & 4.15 & 4.35 \\
Ours (Base)                      & 95.81 & \textbf{86.81} & 94.23 & 92.11   & 88.20 & 91.43 && 1.64  & 2.83    & 10.38 & 2.80    & 4.93 & 4.52 \\
Ours (CBAM)                      & 95.92 & 86.63          & 94.48 & 92.57   & 88.15 & 91.55 && 1.55  & 2.99    & 10.00 & 3.68    & 4.43 & 4.53 \\
Ours (Full)                      & 96.12 & 86.63          & 95.56 & 91.70   & 88.45 & 91.69 && 1.56  & 2.97    & 9.56  & 3.24    & 4.40 & 4.35 \\
\textbf{Ours (Mid)} &
  \textbf{96.12} &
  86.48 &
  \textbf{97.40} &
  \textbf{92.26} &
  \textbf{88.51} &
  \textbf{92.15} &&
  \textbf{1.53} &
  \textbf{2.93} &
  \textbf{6.54} &
  \textbf{2.80} &
  \textbf{4.12} &
  \textbf{3.64} \\
  \bottomrule
\end{tabular}
\end{table}

\begin{table} [!h]
\setlength{\tabcolsep}{5pt}
\footnotesize
\centering
\caption{Quantitative results of proposed methods compared to state-of-the-art methods for BraTS 2020. (Bold numbers are the best results)}
\label{tab8}
\begin{tabular}{lcccclcccc}
\toprule
\multirow{2}{*}{\textbf{Method}} & \multicolumn{4}{l}{\textbf{Dice$\uparrow$}} && \multicolumn{4}{l}{\textbf{HD95$\downarrow$}} \\
 & \multicolumn{1}{l}{ET} & \multicolumn{1}{l}{WT} & \multicolumn{1}{l}{TC} & \multicolumn{1}{l}{mean} && \multicolumn{1}{l}{ET} & \multicolumn{1}{l}{WT} & \multicolumn{1}{l}{TC} & \multicolumn{1}{l}{mean} \\
 \midrule
Myronenko \cite{myronenko3DMRIBrain2019} & 64.77 & 84.31 & 72.61 & 73.90 && 41.35 & 13.85 & 18.57 & 24.59 \\
Wang \cite{wangModalityPairingLearningBrain2021} & 78.70 & 90.80 & 85.60 & 85.03 && 35.01 & 4.71 & 5.70 & 15.14 \\
H2NF-Net \cite{jiaH2NFNetBrainTumor2021} & 78.75 & 91.29 & 85.46 & 85.17 && 26.58 & 4.18 & \textbf{4.97} & 11.91 \\
nnU-Net (Baseline) \cite{isenseeNnUNetBrainTumor2021} & 77.67 & 90.60 & 84.26 & 84.18 && 35.10 & 4.89 & 5.91 & 15.30 \\
nnU-Net (Best) \cite{isenseeNnUNetBrainTumor2021} & 79.85 & 91.18 & 85.71 & 85.58 && 26.41 & 3.71 & 5.64 & 11.92 \\
Ours (Base) & 78.94 & 91.18 & 84.99 & 85.04 && 29.14 & 4.77 & 6.01 & 13.31 \\
Ours (Full) & 79.01 & 91.31 & 85.75 & 85.36 && 26.27 & 4.56 & 5.87 & 12.23 \\
\textbf{Ours (Mid)} & \textbf{79.94} & \textbf{91.68} & \textbf{85.82} & \textbf{85.81} && \textbf{25.22} & \textbf{3.65} & 5.02 & \textbf{11.30}\\
\bottomrule
\end{tabular}
\end{table}

\begin{figure}[h]
\includegraphics[width=\textwidth]{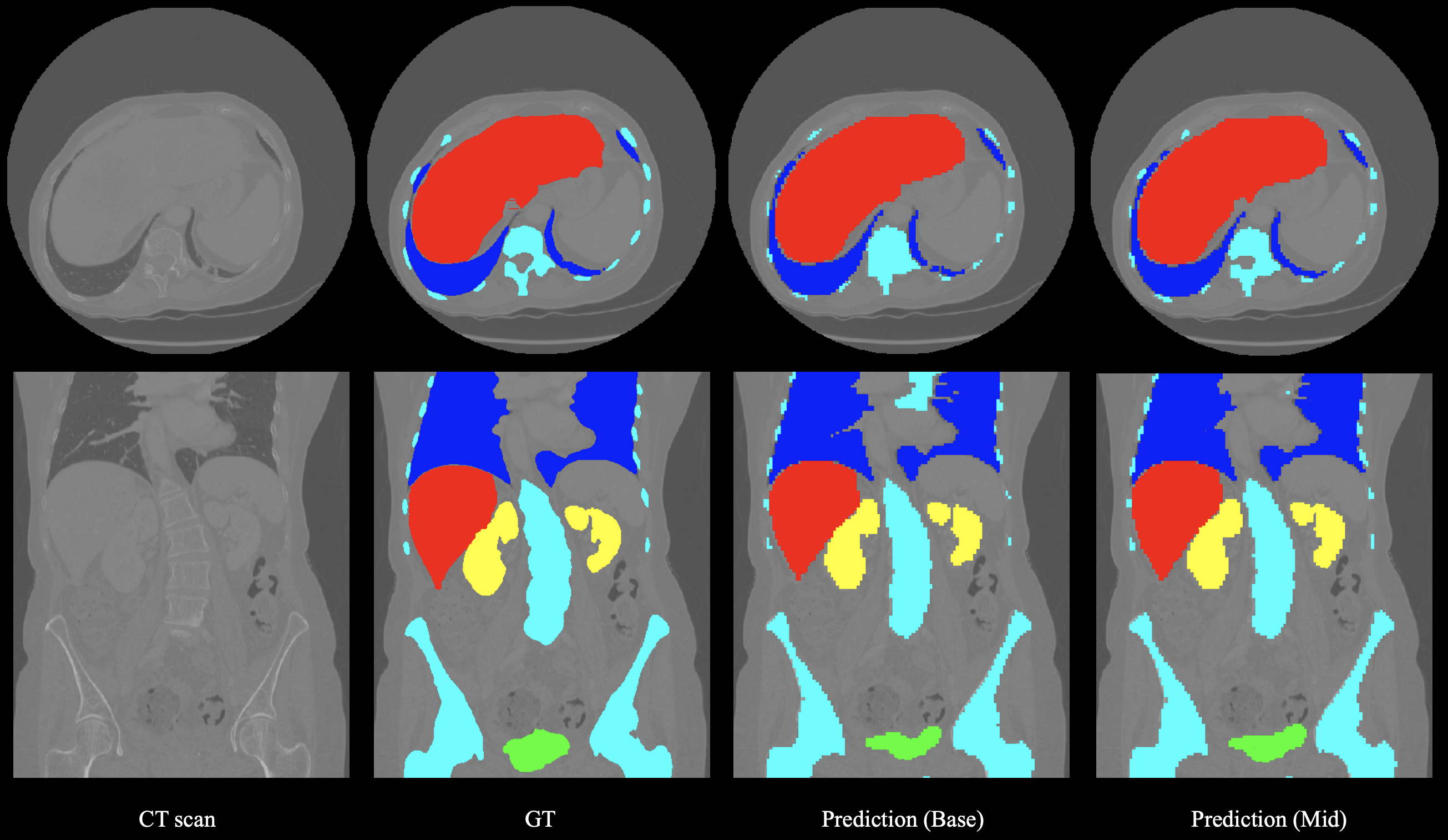}
\caption{Representative visual results of proposed methods for CT-ORG. From left to right: CT scan, ground truth (GT), and predictions. The labels are liver (red), gladder (green), lungs (blue), kidneys (yellow), and bone (cyan).} 
\label{fig3}
\end{figure}

\begin{figure}[h]
\includegraphics[width=\textwidth]{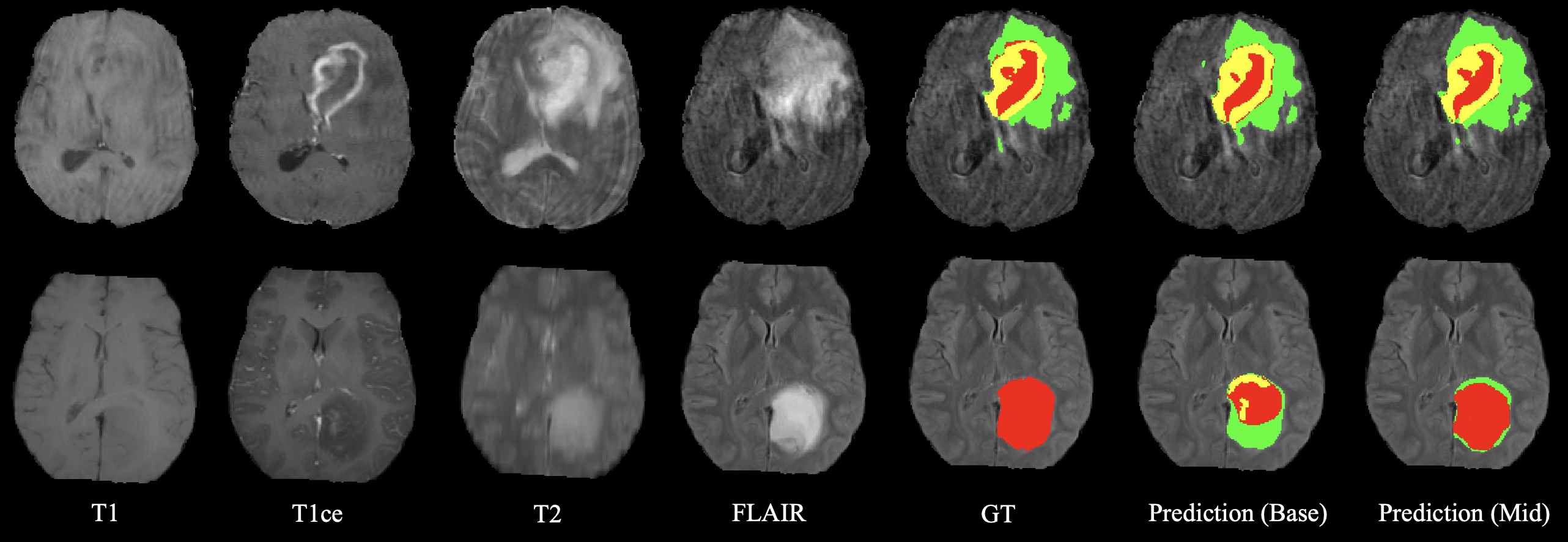}
\caption{Representative visual results of proposed methods for BraTS 2020.From left to right: four MRI modalities, ground truth (GT), and predictions. The labels are enhancing tumor (yellow), edema (green), and necrotic and non-enhancing tumor (red).} 
\label{fig4}
\end{figure}

Quantitative results show that the proposed Mid-type network outperformed other model architectures and all state-of-the-art methods for segmenting all organs and tumor subregions. For multi-organ segmentation, the proposed method achieved the highest Dice score and the lowest HD95 score in all organs, especially the lungs. This might be attributed to the fact that the LK attention module emphasizes the features of the correct organ, thereby reducing distracting and false predictions. For the Dice score, the Mid network was only slightly inferior to the Base network in the segmentation of the bladder. We found that adding any attention mechanism caused a decrease in Dice for bladder segmentation. This may be due to the uneven distribution of attention to fine organs, resulting in a greater concentration of computing power on other organs. For brain tumor segmentation, the proposed method was only slightly lower than other methods in the HD95 score of TC by a tiny 0.05 margin. On the other hand, the Mid network performed very well on ET's HD95 score, also due to the LK attention module adding feature weights to the correct tumor subregions. Representative segmentation results are also compared visually in Figures~\ref{fig3} and~\ref{fig4}, which further proves the effectiveness of the LK attention module. 

\begin{figure}[h]
\includegraphics[width=\textwidth]{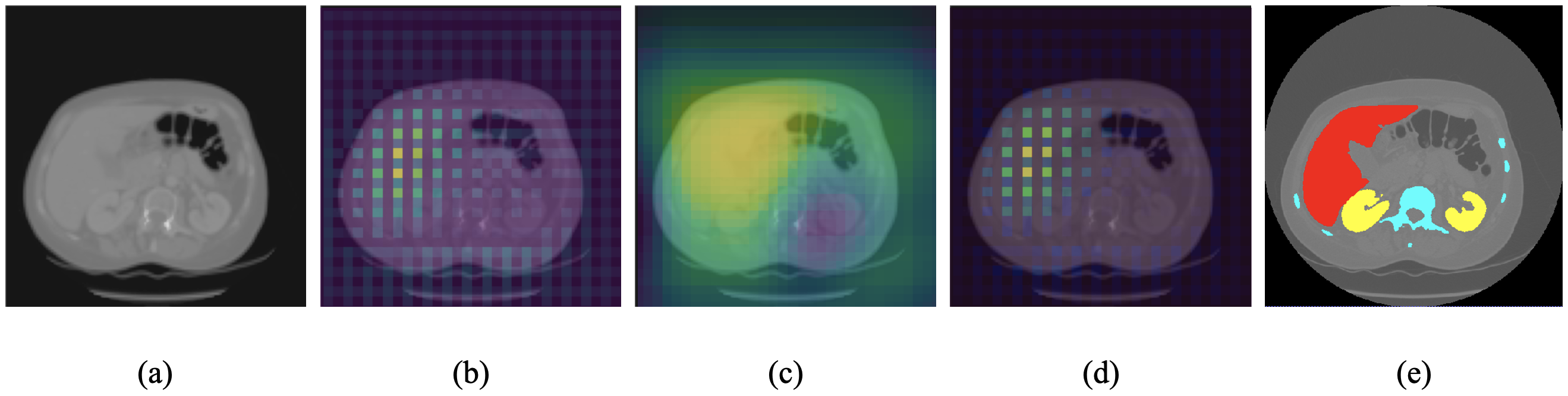}
\caption{A representative visual effect of the LK attention module. (a) The CT scan input. (b) The upsampled feature map at the middle scale of the decoder. (c) The attention map. (d) The feature map after multiplying with the attention map. (e) The GT labels.} 
\label{fig5}
\end{figure}

Comparing the results of the Base and Mid networks, the performance improvement due to the presence of the LK attention module can be seen. Lungs, ET, and TC had more significant improvements in segmentation performance, as shown in Tables~\ref{tab9} and~\ref{tab10}. To verify the metric's increase, we performed a paired t-test, and the p-values are shown in Table 6. The improvements brought by the LK attention module on all segmentation targets were statistically validated, except for bladder and ET. The LK attention module caused a slight decrease in the accuracy of bladder segmentation. On the other hand, since BraTS 2020 set a penalty of Dice = 0 and HD95 = 373.13 for false positives of ET, the paired t-test cannot verify changes in ET. This statistic validates the effectiveness of the adaptive feature selection of the LK attention module, as visualized in Figure~\ref{fig5}.

Furthermore, high-performance deep learning models usually produce incomprehensible results for humans. While these models can yield better efficiencies than humans, it is not easy to express intuitive explanations to justify their findings or to derive additional clinical insights from these computational "black boxes" \cite{yangUnboxBlackboxMedical2022}. Given the importance of explainability in the clinical domain, our proposed LK attention module proved that deep learning models could identify appropriate regions in medical images without overemphasizing unimportant findings. The local explanation furnished directly by the LK attention map (in Figure~\ref{fig5}) argued that there was medical reasoning for the focused parts of the CT scan, which could facilitate clinicians' decision-making.

\begin{table} [!h]
\setlength{\tabcolsep}{1.5pt}
\footnotesize
\centering
\caption{Improvement in quantitative results due to the LK attention module for CT-ORG.}
\label{tab9}
\begin{tabular}{lcccccclcccccc}
\toprule
\multirow{2}{*}{} & \multicolumn{6}{l}{\textbf{Dice$\uparrow$}} && \multicolumn{6}{l}{\textbf{HD95$\downarrow$}} \\
 & \multicolumn{1}{l}{liver} & \multicolumn{1}{l}{bladder} & \multicolumn{1}{l}{lungs} & \multicolumn{1}{l}{kidneys} & \multicolumn{1}{l}{bone} & \multicolumn{1}{l}{mean} && \multicolumn{1}{l}{liver} & \multicolumn{1}{l}{bladder} & \multicolumn{1}{l}{lungs} & \multicolumn{1}{l}{kidneys} & \multicolumn{1}{l}{bone} & \multicolumn{1}{l}{mean} \\
 \midrule
Ours (Base) & 95.81 & 86.81 & 94.23 & 92.11 & 88.20 & 91.43 && 1.64 & 2.83 & 10.38 & 2.80 & 4.93 & 4.52 \\
Ours (Mid) & 96.12 & 86.48 & 97.40 & 92.26 & 88.51 & 92.15 && 1.53 & 2.93 & 6.54 & 2.80 & 4.12 & 3.64 \\
Improvement & 0.31 & -0.33 & 3.17 & 0.15 & 0.31 & 0.72 && -0.12 & 0.10 & -3.84 & 0.00 & -0.82 & -0.88 \\
p-value & 0.020 & 0.031 & 0.040 & 0.026 & 0.051 & 0.030 && 0.021 & 0.025 & 0.025 & 0.059 & 0.015 & 0.028 \\
\bottomrule
\end{tabular}
\end{table}

\begin{table} [!h]
\footnotesize
\centering
\caption{Improvement in quantitative results due to the LK attention module for BraTS 2020.}
\label{tab10}
\begin{tabular}{lcccclcccc}
\toprule
\multirow{2}{*}{} & \multicolumn{4}{l}{\textbf{Dice$\uparrow$}} && \multicolumn{4}{l}{\textbf{HD95$\downarrow$}} \\
 & \multicolumn{1}{l}{ET} & \multicolumn{1}{l}{WT} & \multicolumn{1}{l}{TC} & \multicolumn{1}{l}{mean} && \multicolumn{1}{l}{ET} & \multicolumn{1}{l}{WT} & \multicolumn{1}{l}{TC} & \multicolumn{1}{l}{mean} \\
 \midrule
Ours (Base) & 78.94 & 91.18 & 84.99 & 85.04 && 29.14 & 4.77 & 6.01 & 13.31 \\
Ours (Mid) & 79.94 & 91.68 & 85.82 & 85.81 && 25.22 & 3.65 & 5.02 & 11.30 \\
Improvement & 1.00 & 0.50 & 0.84 & 0.78 && -3.92 & -1.12 & -0.99 & -2.01 \\
p-value & 0.286 & 0.013 & 0.013 & 0.015 && 0.095 & 0.037 & 0.044 & 0.65\\
\bottomrule
\end{tabular}
\end{table}

\subsection{Limitations}

Our method still has some limitations. First, as shown in Figure~\ref{fig3}, the segmentation results' resolution was lower than GT due to resampling. In future work, the resolution of the segmentation mask can be improved by resampling the image to a higher resolution with sliding windows. Moreover, in the second example of Figure~\ref{fig4}, the TC was not accurately segmented, which might be due to the blurring of the T2 modality. This demonstrates the importance of data integrity for the accurate segmentation of medical images. This can be solved by more diverse data acquisition and data augmentation or by training generative networks to synthesize clear images.

\section{Conclusion}\label{sec13}

This paper introduced LK attention for 3D medical image segmentation, which can be easily incorporated into any FCN such as U-Net. The LK attention module combines the advantages of convolution and self-attention, exploits local contextual information, long-range dependencies, spatial and channel adaptation, and uses convolutional decomposition to eliminate the disadvantage of high computational cost. Ablation experiments on the CT-ORG dataset first verified the feasibility of the decomposition of LK convolutions and secondly explored the most efficient deployment design of the LK attention module. The quantitative results of ablation learning indicated that incorporating the LK attention module in the middle of the decoder achieved optimal performance. The Mid-type LK attention-based U-Net achieved state-of-the-art performance on both multi-organ and tumor segmentation. Segmentation results of CT-ORG and BraTS 2020 datasets showed that the LK attention module improved predictions for all organs and tumor subregions except the bladder, especially for lung, ET, and TC.  In addition, the LK attention module was proven to be effective in adaptively selecting important features and suppressing noise, which provided local explanations of model's prediction.

However, some challenges remained. First, the addition of attention caused the scattered computing power for some fine targets such as the bladder. Thus, the LK attention module can be further customized for multi-target segmentation. Second, for large medical images, better sampling or training strategies can be used to further improve the resolution of the segmentation results. Furthermore, since the low quality of the images can significantly reduce the segmentation accuracy, more comprehensive data augmentation strategies and larger training datasets can be considered, or a generative network can be used to synthesize high-quality images.

\backmatter

\bmhead{Acknowledgments}

This study was supported in part by the BHF (TG/18/5/34111, PG/16/78/32402), the ERC IMI (101005122), the H2020 (952172), the MRC (MC/PC/21013), the Royal Society (IEC/NSFC/211235), the Imperial College Undergraduate Research Opportunities Programme (UROP), the NVIDIA Academic Hardware Grant Program, the SABER project supported by Boehringer Ingelheim Ltd, NIHR Imperial Biomedical Research Centre (RDA01), and the UKRI Future Leaders Fellowship (MR/V023799/1). J. Del Ser also acknowledges funding support from the Department of Education of the Basque Government (Consolidated Research Group MATHMODE, IT1456-22).

\section*{Declarations}

\subsection*{Declaration of competing interest}
The authors have no relevant financial or non-financial interests to disclose. 

\subsection*{Data availability statement}
The datasets generated during and/or analysed during the current study are available from the corresponding author on reasonable request. 

\subsection*{CRediT authorship contribution statement}
\textbf{Hao Li}: Conceptualization, Methodology, Software, Validation, Formal analysis, Investigation, Writing - original draft. \textbf{Yang Nan}: Conceptualization, Methodology, Writing - review and editing, Supervision. \textbf{Javier Del Ser}: Writing - review and editing, Supervision. \textbf{Guang Yang}: Conceptualization, Methodology, Writing - review and editing, Supervision, Funding acquisition.


\bibliography{LKA.bib}


\end{document}